\documentclass[12pt]{article}

\begin{document} 
\bibliographystyle{unsrt} 
\centerline{\LARGE Non-monotonic field-dependence }
\centerline{\LARGE of the ZFC magnetization peak}
\centerline{\LARGE in some systems of magnetic nanoparticles}

\vskip 0.2 cm \noindent R. Sappey, E. Vincent, N. Hadacek\\ {\it
Service de Physique de l'Etat  Condens\'e, CEA Saclay,\\ 91191 Gif sur
Yvette Cedex, France \\} \\ F. Chaput, J.P. Boilot, \\ {\it Groupe de
Chimie des Solides, Laboratoire P.M.C., CNRS URA D1254,\\ Ecole
Polytechnique, 91128 Palaiseau, France \\} \\ D. Zins, \\ {\it
Laboratoire de Physico-Chimie, CNRS ER 44,\\ Universit\'e Pierre et
Marie Curie, 4 place Jussieu, 75252 Paris, France}

\def\gam{$\gamma-Fe_2O_3$}

\vskip 0.2 cm \begin{abstract}
We have performed magnetic measurements
on a diluted system of \gam ~nanoparticles ($d\sim 7nm$), and on a
ferritin sample. In both cases, the  ZFC-peak presents a non-monotonic
field dependence, as has already been reported in some experiments,
and discussed as a possible evidence of resonant tunneling. Within
simple assumptions, we derive expressions for the magnetization
obtained in the usual ZFC, FC, TRM procedures. We point out that the
ZFC-peak position is extremely sensitive to the width of the particle
size distribution, and give some numerical estimates of this
effect. We propose to combine the
FC magnetization with a modified TRM measurement, a procedure which allows a 
more direct access to the barrier distribution in a field. The typical
barrier values which are obtained with this method show a monotonic
decrease for increasing fields, as expected from the simple effect of
anisotropy barrier lowering, in contrast with the ZFC results. From
our measurements on \gam ~particles, we show that the width of the
effective barrier distribution is slightly increasing with the field,
an effect which is sufficient for causing the observed initial
increase of the ZFC-peak temperatures. 

\end{abstract}
 \vskip 0.2 cm \noindent P.A.C.S. numbers : 75.50.Tt,
75.45.+j, 75.60.Nt

\noindent {\small E-mail: vincent@spec.saclay.cea.fr}

\pagenumbering{arabic}  \pagestyle{plain} 

\def\gam{$\gamma-Fe_2O_3$}

\section{Introduction}

A rapid characterization of ensembles of small magnetic particles
(like ferrofluids) is very commonly achieved by ``zero-field cooled''
(ZFC) magnetization measurements. The ZFC curve is measured by cooling
the sample in zero field, applying the field at low temperature and
then measuring the magnetization while raising the temperature by
steps.  The ZFC curve peaks at a temperature which is related to a
typical scale of the anisotropy energy barriers in the system; it is
commonly referred to as the ``blocking temperature'' of the sample.
For ZFC curves measured under increasing field amplitudes, the peak is
expected to reflect the lowering of the anisotropy barriers, and hence
should shift towards lower temperatures (as observed e.g. in
\cite{teja1}).

However, in several experiments \cite{luo,awscha,myriam,tejachud}, an
astonishing increase of the ZFC-peak temperature with the field
amplitude has been reported. In the first papers \cite{luo,awscha}, no
explanation was proposed for this apparent barrier increase under the
effect of the applied field. In very recent works on antiferromagnetic
particles of ferritin \cite{myriam,tejachud}, interestingly, the
effect has been discussed as a possible indication of a resonant spin
tunneling phenomenon \cite{chudres}. In brief, if the magnetic moment
of the particles can flip by quantum tunneling through the anisotropy
barrier (a process which should be favored in antiferromagnetic
particles \cite{chudAF}), then the flipping rate should be enhanced by
a resonance effect when the up and down energy levels coincide. In
Mn-12 magnetic molecules, where the energy levels can be well defined,
the resonances have been recently observed for the corresponding
values of the field \cite{mn12ny,mn12barb}. In a system of
size-distributed particles, there can be no coincidence of the various
up and down energy levels in the different particles, except in the
symmetrical situation of zero field. Resonant tunneling  has thus been
suggested to produce an increase of the relaxation rate around zero
field \cite{chudres}, which could (among other evidences, see
\cite{myriam,tejachud}) show up as the observed anomalous increase of
the ZFC-peak temperature for increasing fields.

In the present paper, we want to address the question of the origin of
this anomalous behavior, and to argue in favor of other
characterization procedures than the ZFC measurement. We first present
a series of experiments on a sample of \gam ~particles, which do
indeed exhibit the ZFC anomaly in the $\sim65K$ region, a rather high
temperature range for expecting evidences of quantum effects. Under
some simple approximations, we discuss the expression of the ZFC
magnetization, and point out that the peak temperature is strongly
influenced by the width of the barrier distribution. We propose as a
possible explanation of the anomaly that this width increases under
the influence of increasing field. 

In comparison with the ZFC-peak results, we use another experimental
procedure, which also gives access to a characteristic temperature
depending on the applied field amplitude. This other characteristic
temperature can be expected to be much less sensitive to the width of
the barrier distribution (and even insensitive in an ideal log-normal
case). Our measurements on \gam ~particles indeed show that this
characteristic temperature decreases for increasing fields, without
any anomaly. We also extract from the \mbox{\gam} ~measurements an approximate
width of the barrier distribution, which we find to slightly increase
with field; the effect has the correct order of magnitude for
reproducing the observed ZFC-anomaly. 

The largest part of the present paper (Sect. 3 and 4) is devoted to
the \mbox{\gam} \ sample, which we have studied in more details until
now \cite{eplnous,icm97,romain}. We use these results as an example
for discussing the physical information which can be extracted from
the various experimental procedures.  Finally, in Sect. 5, we apply
the same procedures to a ferritin sample. The
anomaly is found in the ZFC measurements around $3000Oe$ (in agreement 
with the other
works \cite{awscha,myriam,tejachud}), and disappears with
the other procedure, making likely our ``classical'' explanation of the
ZFC-anomaly.

\section{Experimental procedure and samples}

Our first sample consists in small ferrimagnetic particles of \gam \
(maghe\-mite), which have been embedded in a silica matrix obtained by
a  room temperature polymerization process \cite{chaput}. Other
samples of the same batch have recently been used for studying the
features of the magnetic relaxation in the limit of very low
temperatures \cite{eplnous,icm97}.  Here, the particles are diluted to
the very low volume fraction of $f_v=2 \ 10^{-4}$, in order to favor
independent relaxation processes of the particles. In a saturated
sample (all particle moments being aligned, which is far from our
case), the corresponding dipolar field would be of order $1 \ Oe$.

We could not directly observe the \gam ~particles in the TEOS
matrix. However, TEM imaging of the  particles before their
incorporation in silica has been made;  Fig.1 displays the resulting
diameter histogram, which can be tentatively fitted (as is usually
done in the literature) to a log-normal shape

\begin{equation}  \label{lognorm} f(d)={1 \over \sqrt{ 2 \pi} \
\sigma_d \ d} \exp \left(-{\ln^2 {d \over d_0} \over 2 \
\sigma_d^2}\right) \ \ , \end{equation} \noindent yielding $d_0 = 7 \
nm$ and $\sigma_d=0.3$. 

We have performed the magnetization measurements with a commercial
SQUID magnetometer (from Cryogenic Ltd, U.K.). Fig.2 presents example
curves from the \gam ~sample, obtained at a given field amplitude
along various procedures. The ZFC curve is measured as explained
above.  The ``FC" (Field Cooled) curve is obtained by cooling the
sample in the field, and measuring while increasing the
temperature. We have used in addition a less common measurement
procedure, which we denote as ``R-TRM" (Reversed Thermo-Remanent
Magnetization); it consists in cooling the sample in the field,
reversing the field at low temperature, and then measuring upon
increasing the temperature. Compared to the more usual ``TRM"
procedure, in which the field is cut-off instead of being reversed, it
presents the advantage that the field conditions for the initial and
final states of the particle relaxation are identical; the effect of
the field amplitude on the barrier distribution can be studied more
directly, as we argue below.

Our second sample in this study is made of horse-spleen commercial
ferritin (Sigma Chimie). Ferritin is an iron-storage protein; it
consists in a protein shell of outer and inner diameters $12nm$ and
$7.5nm$, which is partially or completely filled with an
antiferromagnetic iron oxide core (maximum of $\sim 5000$ Fe ions per
ferritin molecule) \cite{haggis}. The concentration of our solution is
$100mg/ml$, which again corresponds to a dipolar field of order $1Oe$
(at saturation of the non-compensated moments). As an example of
anti\-ferromagnetic nano\-particles, ferritin is considered a good
candidate for the observation of quantum tunneling of the N\'eel
vector \cite{chudAF}, and has been the subject of numerous studies at
low temperatures these last years (see
\cite{awscha,myriam,tejachud,awschafer,tejafer} and references
therein).

All throughout the paper, we have chosen as a convention to present
the results in terms of magnetic moments, in 
c.g.s. electro\-magnetic units; we have not divided the measured
magnetic moments by the sample volume, which we estimate for the \gam
~particles to $V_{tot}=2.1\,10^{-5} \ cm^3$. For ferritin, we only
know the total mass, which amounts to $8.4\ 10^{-3}g$ of ferritin
particles. Coherently, in the following equations, we do not divide
by integrals over the particle volumes.

\section{ZFC measurements: anomalous field dependence}

We present now the ZFC measurements which we have performed on our
sample of \gam ~particles, for field amplitudes ranging from $1\ {\rm
to}\ 200 \  Oe$ (in this sample, the effective coercive field which
brings the total magnetization to zero after saturation is  $\sim 300 \  Oe$
at 2 K \cite{romain}). The curves are displayed in Fig.3a, and the
peak temperature variation with the field is shown in
Fig.3b. Surprisingly, the peak temperature increases with the field up
to $\sim 80 \ Oe$, before decreasing for larger values as expected. 

The initial increase of a ZFC curve reflects the additive
contributions of larger and larger particles which are deblocked as
the temperature is raised; the maximum is obtained when these
contributions are compensated by the superparamagnetic reduction of
already deblocked moments.  It is therefore clear that the peak
temperature has no simple relation with the peak of the size
distribution.  One may however consider that it is related to some
typical anisotropy barrier; in that case, the effect of an increasing
field amplitude should be to lower the anisotropy barrier, in
contradiction with our result in Fig.3b.

A similar observation has already been reported for magnetite
particles \cite{luo}, and also in  ferritin \cite{awscha}; no
explanation was proposed. Again in ferritin, the phenomenon has
recently been  quoted \cite{myriam,tejachud}, and discussed as a
possible indication of a resonant tunneling process at zero field
\cite{chudres}. In our present sample, the temperature range of the
ZFC-peak ($\sim 65 \ K$) does not favor an explanation of quantum
origin. In the following, we write in more details the $M_{ZFC}$
expression under simple assumptions, and propose a semi-quantitative
explanation of a non-monotonic behavior of the peak temperature in
terms of the field influence on the barrier distribution.

The ZFC data being taken in a field $H$, deblocking of particles with
anisotropy barrier $U(H)$ occurs at a temperature $T_b$ such that the
typical time for crossing the barrier $U(H)$ is equal to the
measurement time $t_m \sim 100 \ s$, namely 

\begin{equation}  \label{Tb} k_BT_b={U(H) \over \ln t_m/\tau_0}
\end{equation} where the attempt time $\tau_0$ is of order
$10^{-10}s$, giving $\ln t_m/\tau_0 \simeq 28$. We assume that the
anisotropy barrier $U$ of a particle is proportional to its volume
$V$; in zero field, $U=KV$, where $K$ is the energy density for
uniaxial  anisotropy (from other measurements, $K\simeq 6\,10^5 \
erg/cm^3$ \cite{romain}). In the general case of random orientations
of the easy axes of the particles, the question of the field
dependence $U(H)$ of the anisotropy barriers cannot be solved
analytically (approximations are discussed in \cite{dormann}).  If the
easy axes are parallel to the field, in contrast, it is
straightforward to derive exactly \begin{equation}
\label{U(H)} U(H)=KV(1-{H\over H_c})^\alpha \end{equation} with
$\alpha=2$. $H_c$ is the coercive field, at which the given barrier
vanishes. In \cite{sampaio}, it has been observed that the disorder of
the easy axes orientations yields a distribution of the $H_c$
values. We restrict ourselves to simply considering that we can
approximate the orientational disorder by Eq.\ref{U(H)} with
$\alpha=1.5$ instead of $\alpha=2$ \cite{victora}, keeping the same
$H_c$ for all particles.

At a given temperature $T$, the magnetization $M_{ZFC}$ is the sum of
the super\-paramagnetic contributions of the particles for which
$T_b<T$, or in other words of volume smaller than a blocking value
$V_b$ such that \begin{equation}         \label{Vb} V_b(T,H)= {k_BT
\ln t_m/\tau_0 \over K(1-H/H_c)^\alpha }.  \end{equation} For the sake
of simplicity, we approximate here the super\-paramagnetic behavior by
an $1/T$ Curie shape, and do not include a temperature dependence of
the saturated magnetization $M_s$. We do not expect these
approximations to significantly affect the present discussion (see
more detailed analysis in \cite{romain}).

Within this framework, $M_{ZFC}$ reads

\begin{equation}    \label{ZFC}
 M_{ZFC} (T) =  M_{r}(H) \ \ + \ \ { M_{s}^2 \over 3\ k_B\ T} \ H \
\int_0^{V_b(T,H) }\ f(V) \ V^2  dV \ \ .  \end{equation} \noindent
where $M_r$ stands for the reversible contribution which is due to the
canting of the moments from the easy axes towards the field
direction. This term equals $M_r=M_s^2 \ V_{tot}\ H\ /\ 3\ K$ in the
$T=0$ limit; at non-zero temperatures, it is a correction to the main
term which accounts for the fact that the moments are not exactly
lying along the easy axes. As is usually done, we neglect it in the
present  discussion of the ZFC-peak; we show below that this term
disappears to first order in some other quantities.

Firstly, one sees in Eq. \ref{ZFC} that the temperature dependence of
$M_{ZFC}$ occurs (at least) via $V_b(T,H)$ and the Curie term. The
temperature derivative cannot be written in simple terms, and there is
no explicit expression of the peak temperature (which, however, obeys
a simple first-order differential equation \cite{bookecjt}).
Secondly, the $f(V)$ distribution is here involved through a $V^2f(V)$
contribution, which clearly emphasizes the effect of the largest
particles; the sensitivity of $M_{ZFC}$ to the 
 standard deviation $\sigma_v=3\sigma_d$ is stronger than that of
other quantities which involve lower powers of V, like the one that we
propose below.

In order to quantitatively estimate the sensitivity of $M_{ZFC}$  to
$\sigma_v$, we have performed  numerical calculations of Eq.\ref{ZFC},
which are shown in Fig.4a. The $K$ and $V_0$  parameters have been
adjusted to the values of the experiment; in this elementary
calculation, due to the various approximations, the shape of the ZFC
curves is not completely realistic \cite{romain}. However, one sees
clearly in Fig.4a that the ZFC-peaks shifts extremely rapidly towards
higher temperatures when $\sigma_v$ is increased. In Fig.4b, we
present the ratio of the ZFC-peak temperature to the blocking
temperature for the typical volume $V_0$. For our sample
($\sigma_v\sim 0.9$), the calculation yields a ratio of 4.4
(neglecting the $M_s(T)$ variation should produce a slight
overestimate). In most cases found in the literature, the standard
deviation of the volume distribution is of this same order of
magnitude; the particle volume which is commonly deduced from the
ZFC-peak must therefore be divided by a non-negligible factor before
being compared with $V_0$.

In our opinion, the result in Fig.4b opens the way to a possible
explanation of the $T_b(H)$ increase at low fields, which could be due
to a slight enlargment of the barrier distribution under the influence
of the field. A simple reason for that can be the disorder of
orientations. For randomly oriented particles of a unique size, the
applied field lowers differently the barriers with respect to their
orientation, thus enlarging the barrier distribution. One may also imagine 
that, 
in relation with the defects of a particle, an increasing field 
results in different  
coupling energies of the field to  various parts of the particle, thus 
yielding several different energy barriers.
Whatever its origin, which remains an open question, an
enlargment of the barrier distribution can indeed be found in the analysis of 
 our R-TRM data (see below).

\section{Other measurement procedures for probing the barrier
distribution}

A TRM measurement corresponds to the inverse field-history of the ZFC
procedure; the sample is cooled in the field, the field is cut at low
temperature, and deblocking is measured for increasing temperatures in
zero field.  Keeping the same assumptions as above, the TRM can be
written as the sum of the moments which are still blocked in the
field-cooled state:

\begin{equation}    \label{TRM1}
 M_{TRM} (T) =  { M_{s}^2 \over 3\ k_B } \ H \
\int_{V_b(T,0)}^{\infty}\  {f(V) \ V^2 \over T_b(V,H)} dV \ \ .
\end{equation} Contrary to the ZFC case, no $M_r$ term appears, and
now the $1/T$ term is replaced by $1/T_b$,  since each particle has
kept a magnetization which is equal to the super\-para\-magnetic value
at the blocking temperature $T_b(V,H)$. $T_b$ is obtained from
Eqs. \ref{Tb},\ref{U(H)}, where $t_m$ now corresponds to the time
scale $\tau_c$ of blocking during the field-cooling process. An
estimate of $\tau_c$ can be obtained from the cooling rate $v_c=dT/dt$
($\simeq 0.04K/s$). As the temperature decreases, the Arrhenius
relaxation time $\tau$ for a given barrier abruptly increases, and
freezing occurs when  $\partial \tau (t) /\partial t\sim 1$.  One
finds that $\tau_c$ satisfies \begin{equation} \label{vc} \tau_c
\ln^2{\tau_c \over \tau_0}=-{U \over k_B v_c} \ \ , \end{equation}
which yields $\tau_c \sim 30\ s\sim t_m$ for $U=KV_0$; the $\ln
t_m/\tau_0$ term which is involved in $T_b$ for the TRM procedure is
almost the same as above. Replacing now  $T_b(V,H)$ in Eq.\ref{TRM1},
we obtain: \begin{equation}    \label{TRM2}
 M_{TRM} (T) =  { M_{s}^2 \ln t_m/\tau_0 \over 3 K (1-H/H_c)^\alpha} \
H \ \int_{V_b(T,0)}^\infty \ f(V) \ V \ dV \ \ .  \end{equation}

The only temperature dependence of the TRM occurs in the lower bound
$V_b(T,0)$ of the integral; this allows us to take very simply the
temperature derivative of $M_{TRM}$ \cite{chantrell}, which reads
\begin{equation}  \label{TRM3} {\partial M_{TRM} \over \partial
T}=-{M_s^2\ k_B\ H\over 3\ K^2} {\ln^2 t_m/\tau_0 \over
(1-H/H_c)^\alpha} V_b(T,0)f(V_b(T,0)) \ \ .  \end{equation}

Thus, the TRM derivative gives a direct access to the quantity
$Vf(V)$; if $f(V)$ is log-normal, then $Vf(V)$ peaks at $V=V_0$,
independently of the width of the distribution. This makes a crucial
difference with the ZFC case, for which the peak rapidly shifts  as
$\sigma_v$ increases. However, the blocking volume $V_b(T,0)$ which is
involved in $\partial M_{TRM}/ \partial T$ is the blocking volume in
zero field, because the measurement is performed in zero field. The
effect of the field amplitude only appears through a multiplicative
factor in Eq.\ref{TRM3}; in other words, $\partial M_{TRM} / \partial
T$ does not give access to the field-modulated barrier distribution.

This is our motivation for using another experimental
procedure, which allows the study of the effect of the field amplitude
on the barrier distribution. We have performed a series of
``Reversed-TRM'' measurements (R-TRM) for various field values; after
field-cooling in $+H$, the field is reversed to $-H$ at low
temperature, and the magnetization is measured while increasing the
temperature. An example of such a curve has been given in
Fig.2. Within the same framework as above, the magnetization
$M_{R-TRM}$ at a given temperature $T$ can be written as the sum of
the contributions of the smaller particles, already deblocked at $T$
in $-H$, plus that of the larger ones, still blocked in the $+H$
field-cooled state; again using the $T_b(V,H)$ expression for the
blocked term, one obtains \begin{eqnarray} \label{RTRM} M_{R-TRM}
(T,H) & = & M_r(-H)+{ M_{s}^2 H\over 3}  \left[ -{1 \over k_B T}
\int_0^{ V_b(T,H) }\ f(V) \ V^2 \ dV \ \ \right. \nonumber \\ & &
\left. +{\ln (t_m/\tau_0) \over K(1-H/H_c)^\alpha}   \int_{
V_b(T,H)}^\infty \ f(V) \ V \ dV  \right] \ \ .  \end{eqnarray} 

This expression looks rather complicated; but it is almost the same as
that of the field-cooled magnetization $M_{FC}$, up to the respective
signs of the super\-para\-magnetic contributions  (also, the
reversible parts $M_r$ are just of opposite sign). In a $+H$ field,
$M_{FC}$ reads:

\begin{eqnarray}    \label{FC} M_{FC} (T,H) & = & M_r(+H)+{ M_{s}^2
H\over 3}  \left[ {1 \over k_B T} \int_0^{ V_b(T,H) }\ f(V) \ V^2 \ dV
\ \ \right. \nonumber \\ & & +{\ln (t_m/\tau_0) \over
K(1-H/H_c)^\alpha}   \left. \int_{ V_b(T,H)}^\infty \ f(V) \ V \ dV
\right] \ \ .  \end{eqnarray} 

The idea is to consider the sum $M_{R-TRM}+M_{FC}$ of both
magnetizations, and thus get rid of the super\-paramagnetic
contribution (and of $M_r$), which presents the most intricate
temperature dependence: \begin{equation}  \label{RTRM+FC} M_{R-TRM}
(T,H)+M_{FC} (T,H)=2\ { M_{s}^2 H\over 3}  {\ln (t_m/\tau_0) \over
K(1-H/H_c)^\alpha}  \int_{V_b(T,H) }^{\infty} \ f(V) \ V \ dV  \ \ .
\end{equation}

As in the TRM case (Eqs. \ref{TRM1},\ref{TRM2}), the temperature
derivative can easily be taken:

\begin{equation} \label{dsum} {\partial (M_{R-TRM}+M_{FC}) \over
\partial T}=-2\ {M_s^2\ H\ k_B\ \ln^2 t_m/\tau_0 \over 3K^2(1-H/H_c)^{
2 \alpha}} \ V_b(T,H)f(V_b(T,H)) \ \ .  \end{equation} \noindent In
this quantity, the blocking volume  corresponds to blocking in a field
$H$, a quantity which was not involved in  simple TRM
measurements. Using our R-TRM and FC measurements, we have estimated
the derivatives Eq.\ref{dsum} for our $1-200\ Oe$ measurement fields;
the resulting curves are displayed in Fig.5a. If the $f(V)$
distribution is log-normal, then $Vf(V)$ is a simple gaussian of $\ln
V/V_0$, which peaks at $V_0$ whatever the distribution width. One may
therefore argue that the peak of this quantity in different fields
corresponds to the same objects. Obviously, the assumption of a
log-normal $f(V)$ remains questionable (see below); however, within
this assumption which is the most commonly used, our procedure allows
a clearly  more direct characterization of the barrier distribution
than the ZFC measurement.

The peak temperatures of Fig.5a are plotted versus H in Fig.5b, which
can be compared with the  ZFC data in Fig.3b . The peak temperatures
monotonically decrease with increasing field, whereas the ZFC results
were exhibiting a striking non-monotonic behavior. The peak
temperatures can be fitted to the expected field-dependence
Eq.\ref{U(H)}; fixing $\alpha=1.5$ \cite{victora} and $V_0=180 \ nm^3$
from TEM (Fig.1), we obtain $H_c\simeq 250 \ Oe$ and $K=6.4\,10^{5} \
erg/cm^3$, in good agreement with other estimates \cite{romain}.

Another combination of R-TRM and FC data can be used for checking the
overall coherence of our data and analysis. According to
Eqs. \ref{ZFC},\ref{RTRM} and \ref{FC}, the three kinds of experiments
are related:

\begin{equation} \label{total} M_{ZFC}={1 \over 2}(M_{FC}-M_{R-TRM}) \
\ , \end{equation} or, in other words, given two of the measurements,
the third one can be deduced. Eq.\ref{total} is thus the generalization
to the situation of a non-negligible field
of the well-known relation \mbox{$M_{ZFC}=M_{FC}-M_{TRM}$.} 
Following a remark by D. Fiorani, we note that Eq.\ref{total} 
allows the reader who
prefers to avoid the R-TRM measurements to use, in place of the sum 
$M_{R-TRM}+M_{FC}$, the equivalent quantity $2(M_{FC}-M_{ZFC})$.   

We have checked the validity of Eq.\ref{total} with our \gam ~data.
Fig.6a compares the measured ZFC magnetizations (symbols) with
the ones which are obtained by combining the FC and R-TRM measurements
through
Eq.\ref{total}. They are in rather good agreement, except a slight
amplitude difference in the vicinity of the peak for the lower field
curves. In Fig.6b, we compare the field variation of the ZFC-peaks
obtained in both direct and indirect way; they are fully compatible
within the errors bars, and in particular the non-monotonic behavior
is found in both cases, whereas it does not show up in the FC+R-TRM
analysis of Fig.5b.

The fact that the anomalous behavior of the ZFC-peak does not appear
in a  (FC+R-TRM) measurement, which is less sensitive to the $f(V)$
width, prompts us to propose that the initial increase of the ZFC-peak
with increasing field be related to an increase of the distribution
width. This effect can be searched in the $V_bf(V_b)$ data which were
presented in Fig.5a; in Fig.7, we present differently this same data,
in a way which favors the comparison of the various curves. If
$f(V_b)$ is log-normal, all $V_bf(V_b)$ curves are simple gaussians of
$\ln T$; their peak temperature corresponds to blocking $V_0$ in a
field $H$, that is the peak temperatures are deduced from each other
by a multiplicative factor (which is the effect of the field on the
anisotropy barrier).  In Fig.7, the data is presented as a function of
$\ln T$, and the peaks are superposed by a T-affinity; also, for
clarity, the peak amplitudes have been normalized to one. 

A slight but systematic asymmetry of the curves can be noted; they are
a little bit more spread out on the low-T side. The derivative
estimate of the first points can be less accurate; apart from that
difficulty, the effect suggests that the log-normal approximation is
not completely correct. This may indicate a difference between the
geometrical sizes which are seen by TEM and the effective magnetic
sizes. However, the accuracy with which the size histogram of Fig.1
suggests a log-normal shape is less than that of Fig.7. 
 The universal success of the log-normal shape for particle size
distributions could be more related to practical reasons than really
scientifically grounded. 

Even slightly asymmetric, the curves in Fig.7 show that the width of
the effective distribution increases for increasing field. Within the
present assumptions, we do not intend  to reproduce in details  the
observed ZFC-peak temperature variation, but we can roughly quantify
the effect.  For example, when $H$ goes from 1 to 50 Oe, the
approximative $\sigma_v$ which can be read in Fig.7 increases from 0.8
to 1.1 . For $H_c=250 Oe$ as obtained above, and using Eq.\ref{U(H)}
with $\alpha=1.5$ for the field influence on the barriers, we have
computed the corresponding ZFC curves; the curve with ($H=50Oe,\
\sigma=1.1$) peaks at a 1.3 times higher temperature than the one with
($H=1 Oe,\ \sigma=0.8$).  Hence, for increasing field, the observed
distribution enlargment is enough for producing an increase of the
ZFC-peak temperature, despite the lowering of the barriers. 

\section{Ferritin results}

In ferritin, a non-monotonic variation of the ZFC-peak,
 together with other particular features of the magnetization
relaxation, 
 has  been  discussed
 in terms of resonant tunneling at zero field \cite{myriam,tejachud}.
A ``pinch" of the hysteresis loop is observed around $H=0$
\cite{myriam,tejachud}; viscosity data can be interpreted as showing
an anomaly \cite{tejachud} (not yet clear in \cite{myriam}), but this
latter point still raises the question of a relevant normalization for
the comparison of viscosity data at various fields, which is not yet
completely solved \cite{myriam},\cite{lluis}.  The observation of
resonant tunneling is more plausible in ferritin than in the \gam
~particles, because of the antiferromagnetic character of the
particles, which makes their resultant moment smaller ($\sim50$ iron
moments); the energy level spacing is thus larger, making wider the
field range around zero where the effect can be visible
\cite{chudres}.  Prompted by discussions with some of the authors of
\cite{myriam} and \cite{tejachud}, we have measured a commercial
ferritin sample and applied the same analysis as above for \gam \
particles.

We have performed the measurements for fields ranging from 50 to 6000
Oe. The ZFC curves are shown in Fig.8a, together with the field
dependence of the peaks in Fig.8b. Here again a non-monotonic
variation is found, in agreement with previous works
\cite{awscha,myriam,tejachud}.  Following the procedure of Sect.4, we
have also measured the FC and R-TRM curves at the same fields, and
estimated the temperature derivative of the sum, which is shown in
Fig.9a (peak values in Fig.9b). The result is qualitatively similar to
the case of the \gam ~particles.  
In the region of $\sim3000Oe$ where
the ZFC-peak data show a clear maximum, the peak values of the
derivative monotonically decrease for increasing field.
An anomalous behavior still remains possible within the error bars
below $1000Oe$, but it is located far below the anomaly which is seen in the 
ZFC-results, and more accurate data would be needed for
discussing this point. 
 Thus, on both
samples that we have studied, the same non-monotonic behavior
is obtained from the ZFC peak-temperatures, and the anomaly is not confirmed 
in the
other procedure. The analysis of $\partial (M_{FC}+M_{R-TRM})/\partial
T$ seems therefore able to provide a physical information which is of
much more direct interpretation than that extracted from ZFC
measurements.

\section{Conclusions}

In this paper, we have discussed the physical interpretation of
standard magnetic measurement procedures in systems of nano\-metric
magnetic particles, on the basis of experiments performed with two
very different samples. One is made of ferri\-magnetic particles
(\gam), highly diluted, with a ZFC-peak temperature of $\sim65K$, and
the other of anti\-ferro\-magnetic particles of ferritin, less diluted
but with much lower magnetic moment, with a ZFC peak in the $10-15K$
range. 

In both samples, the ZFC-peak temperature is found to initially
increase with field, at variance with the common sense expectation of
an anistropy barrier lowering due to the field. From a very simple
description of the blocking and deblocking processes, we recall that
the ZFC-peak temperature is not simply related to the typical volume
of the distribution $f(V)$; it is influenced by the $1/T$ behavior of
the deblocked particles, and involves a $V^2f(V)$ term which enhances
the contribution of the larger volumes. The ZFC curve is thus
extremely sensitive to the distribution width; the peak rapidly shifts
to higher temperatures when the width increases, an effect that we
have quantified under simple approximations.

We propose to understand the ZFC anomaly at the light of another
experimental procedure. As a first example, the temperature variation
of the TRM, which is measured in zero field, does not involve the
$1/T$ super\-paramagnetic contribution, and contains a $Vf(V)$ term
which yields a weaker sensitivity to the distribution width. But the
TRM does not bring informations about the effective distribution of
anisotropy barriers in a field. This point can be studied using a
Reversed-TRM procedure, in which the field is reversed to its opposite
value at low temperature. In the sum of the FC magnetization and the
R-TRM, the $1/T$ term is eliminated (together with the reversible
magnetization), and $f(V)$ comes in through $Vf(V)$ (weak sensitivity
to the width), in which $V$ now stands for the volume which is
deblocked in the field, thence the access to the field-modulated
barrier distribution. Note that one may also use the 
equivalent combination $M_{FC}-M_{ZFC}$, which 
presents the same property.

 The temperature derivative $\partial (M_{FC}+M_{R-TRM})/\partial T$
of this sum is proportional to $Vf(V)$, which peaks to a typical
volume in the distribution, and our point is the following: for
different experiments with various field amplitudes, the magnetic
objects which correspond to the peak value remain
 almost the same (exactly the same  in the log-normal case), which is
far from being the case for ZFC measurements.  Indeed, our
measurements on both samples show that the peak of $\partial
(M_{FC}+M_{R-TRM})/\partial T$  decreases for increasing field, in
contrast with the peak of the ZFC curves. 

The effect of the field on the distribution of anisotropy barriers is
not easily described in details \cite{dormann}, mainly for two
reasons. On the one hand, for random orientations of the particle easy
axes, there is no general analytical treatment of the problem. On the
other hand, the usual assumptions which are commonly made for
describing systems of small particles might become less applicable in
the presence of higher fields (is each particle a single fixed
macro-moment, or do some parts couple selectively to the field~?
are the particles relaxing independently, or do they become influenced
by the field of their neighbors~?). On the \gam ~sample, which we have
studied in more details than the ferritin,  the examination of the
measured $\partial (M_{FC}+M_{R-TRM})/\partial T$ shows that, for
increasing field, $Vf(V)$ naturally peaks to lower values, but also
becomes wider, as already expected from the only effect of
orientational disorder. The observed effect has the correct order of magnitude
for compensating the barrier decrease at low fields, and hence for
producing the observed anomalous increase of the ZFC peak
temperature. We therefore consider that the non-monotonic variation of
the ZFC-peak temperature is related to an enlargment of the effective
barrier distribution under the influence of the field; no anomaly is
found using the other procedure.

There has been these last years a renewed interest for the
low-temperature dynamics of systems of small particles, motivated by a
search for quantum tunneling phenomena in these quasi-macroscopic
objects \cite{chudAF}. Evidencing the quantum effects from viscosity
measurements is hindered by the lack of knowledge of the effective
barrier distribution, which modulates the temperature variation of the
measured relaxation rates \cite{eplnous,sampaio}. Very recently,
observations of the non-monotonic field dependence of the ZFC-peak
temperature in ferritin \cite{myriam,tejachud} have been discussed in
terms of possible resonant tunneling effects in zero field
\cite{chudres}. This has prompted us to extend the present work,
mainly centered on \gam ~particles, to a ferritin sample. It appears
that the same ``classical" explanation of the ZFC anomaly should work
in both cases. This conclusion does not concern other possible
evidences of the resonant tunneling effects in ferritin, like e.g. the
 pinched hysteresis cycles \cite{myriam,tejachud}.   Here
again, as is the case for viscosity, it appears that the barrier
distribution plays a non-negligible role, and that the choice of
physically meaningful quantities for characterizing the
low-temperature dynamics of magnetic nanoparticle systems remains a
delicate matter.

\vskip 1.5 truecm We want to thank E.M. Chudnovsky for numerous
stimulating discussions all along this work, and D. Fiorani for a 
useful suggestion.

\newpage

\pagestyle{empty} \noindent{\Large \bf Figure captions}

\vskip 0.5 truecm \underline {Figure 1:} Histogram of the \gam
~particle diameters, as observed in TEM imaging (symbols). 454
particles have been sampled. The dotted line is a fit to a log-normal
distribution, with $d_0=7.05nm$ and $\sigma_d=0.32$. 

\vskip 0.5 truecm \underline {Figure 2:} Example magnetization curves
from the \gam ~particles ($H=80Oe$), obtained following the different
experimental procedures: Field-Cooled, Zero-Field Cooled, and Reversed
Thermo-Remanent Magnetization (the R-TRM curve has been multiplied by
$-1$ in the figure). 

\vskip 0.5 truecm \underline {Figure 3a:} Measured ZFC magnetizations
on the \gam ~sample, normalized to the field amplitude. From top to
bottom, the field values are 1, 10, 20, 50, 80, 110, 150 and 200 Oe.
 
\vskip 0.5 truecm \underline {Figure 3b:}  Peak temperatures of the
measured ZFC magnetization curves for \gam .

\vskip 0.5 truecm \underline {Figure 4a:} Calculated ZFC curves, using
a log-normal volume distribution, for various values of the standard
deviation $\sigma_v$.

\vskip 0.5 truecm \underline {Figure 4b:} Ratio of the calculated
ZFC-peak temperatures to the blocking temperature corresponding to
$V_0$ (reference volume of the log-normal distribution), for different
values of the standard deviation $\sigma_v$.
 
\vskip 0.5 truecm \underline {Figure 5a:} Temperature derivative of
the sum of the measured magnetizations $M_{FC}+M_{R-TRM}$, divided by
the field amplitude, for different fields (\gam ~sample).
   
\vskip 0.5 truecm \underline {Figure 5b:} Peak temperatures of the
curves in Fig.5a, for different fields; these temperatures do not show
the non-monotonic behavior which is found using the ZFC-peaks.
  
\vskip 0.5 truecm \underline {Figure 6a:} Comparison for the \gam
~sample of the measured ZFC magnetizations (symbols) with the
combination of  measured magnetizations $(M_{FC}-M_{R_TRM})/2$ (solid
lines), showing the consistency of the data and of our description
(the magnetizations are normalized to the field amplitude). The field
values are the same as in Fig.3a.   

\vskip 0.5 truecm \underline {Figure 6b:} Comparison of the peak
temperatures of the measured ZFC curves (full circles) with the peak
temperatures of the combination $(M_{FC}-M_{R-TRM})/2$ of other
measured magnetizations (open squares).
  
\vskip 0.5 truecm \underline {Figure 7:} Temperature derivative
(normalized to the peak amplitude) of the combination
$(M_{FC}+M_{R-TRM})$ of measured magnetizations, as a function of the
neperian logarithm of the temperature (normalized to the peak
position), for the \gam ~sample.

\vskip 0.5 truecm \underline {Figure 8a:} Measured ZFC magnetizations
on the ferritin ~sample, normalized to the field amplitude. From top
to bottom, the field values are 50, 200, 600, 1000, 2000, 3000, 4500
and 6000 Oe.
 
\vskip 0.5 truecm \underline {Figure 8b:}  Peak temperatures of the
measured ZFC magnetization curves for ferritin (more data than in Fig.8a).

\vskip 0.5 truecm \underline {Figure 9a:} Temperature derivative of
the sum of the measured magnetizations $M_{FC}+M_{R-TRM}$, divided by
the field amplitude, for different fields (ferritin ~sample).
   
\vskip 0.5 truecm \underline {Figure 9b:} Peak temperatures of the
curves in Fig.9a (ferritin sample, more data than in Fig.9a), for different 
fields, which do not
confirm the non-monotonic behavior observed for the ZFC-peak.

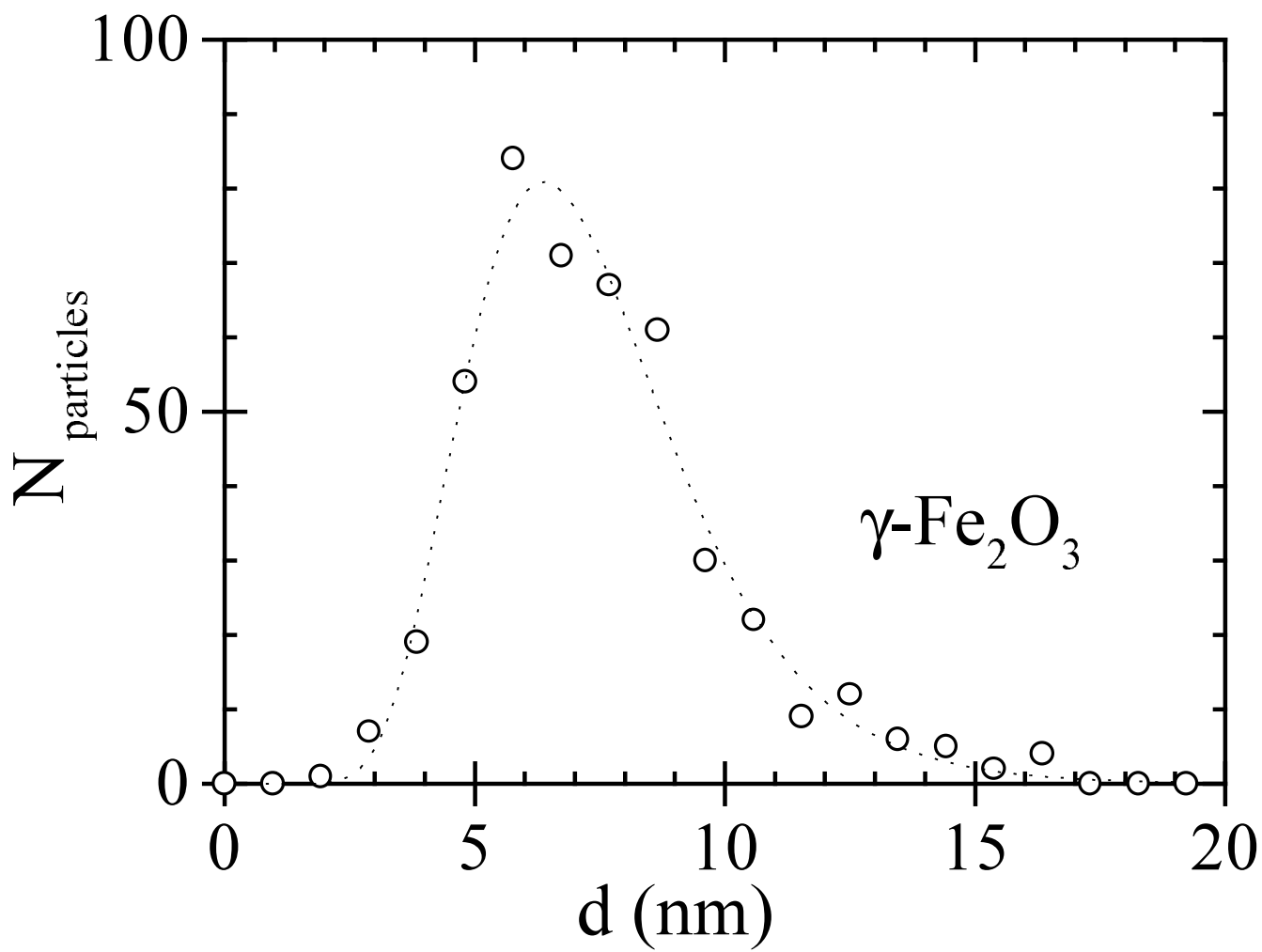
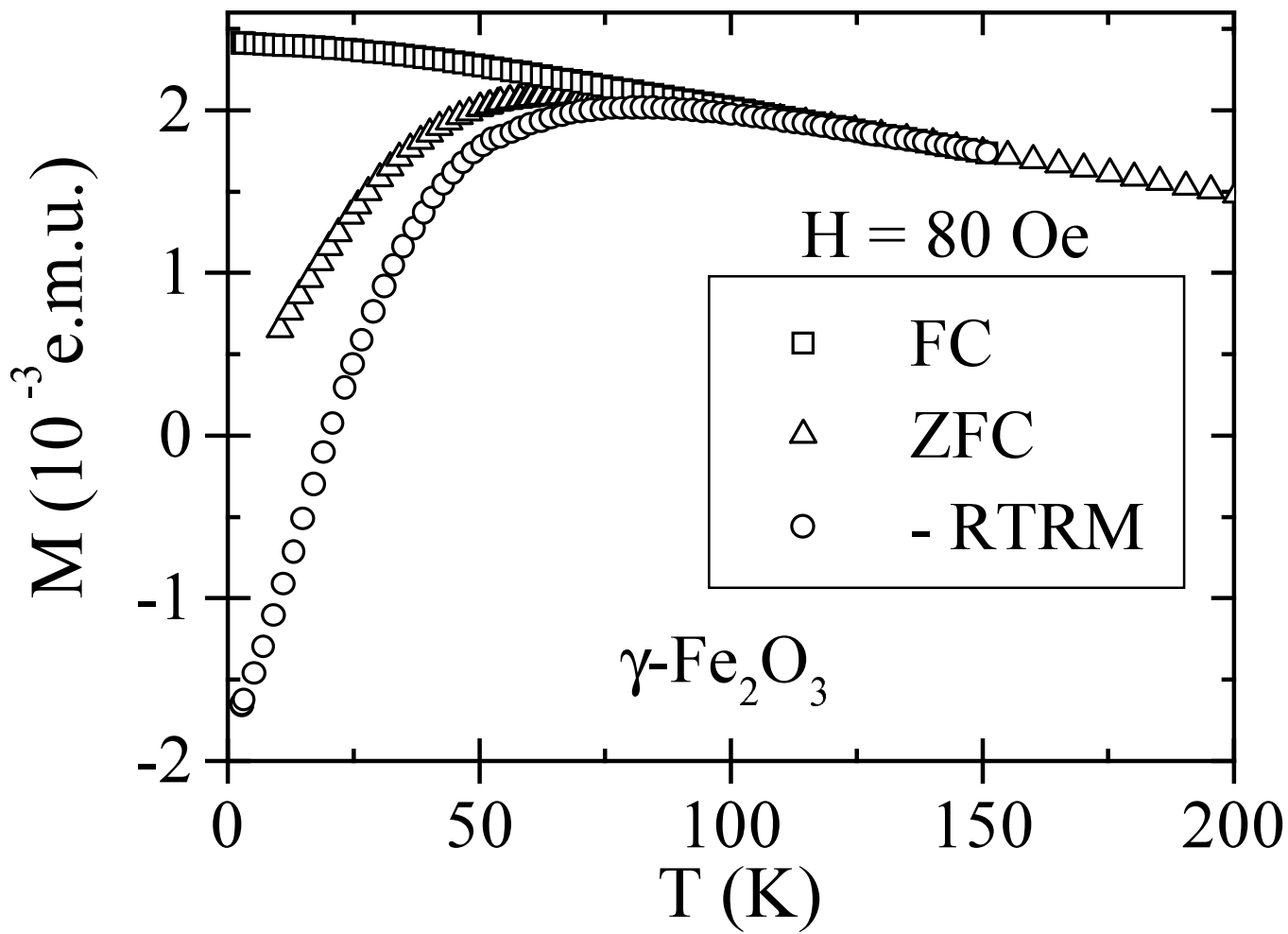
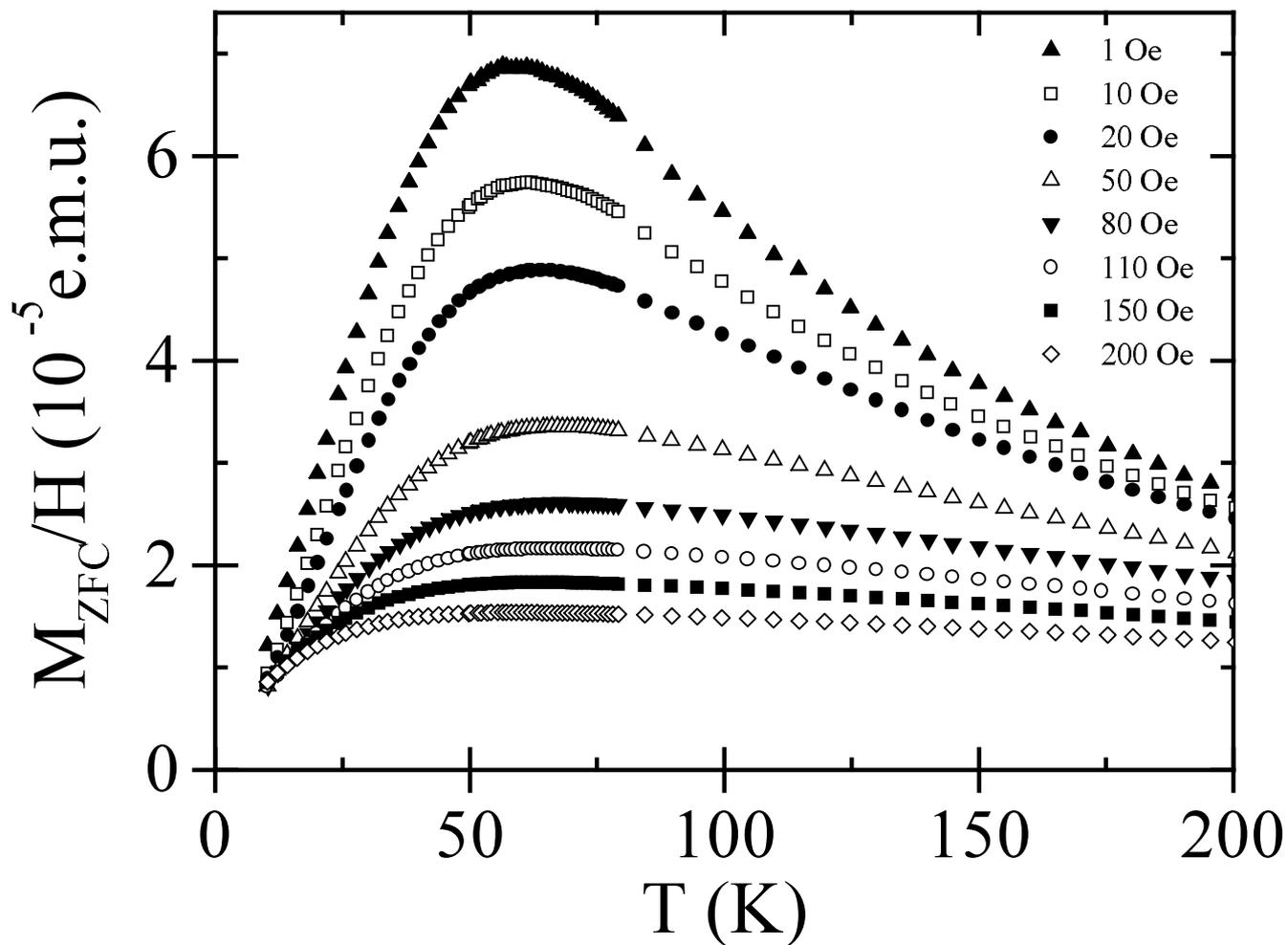
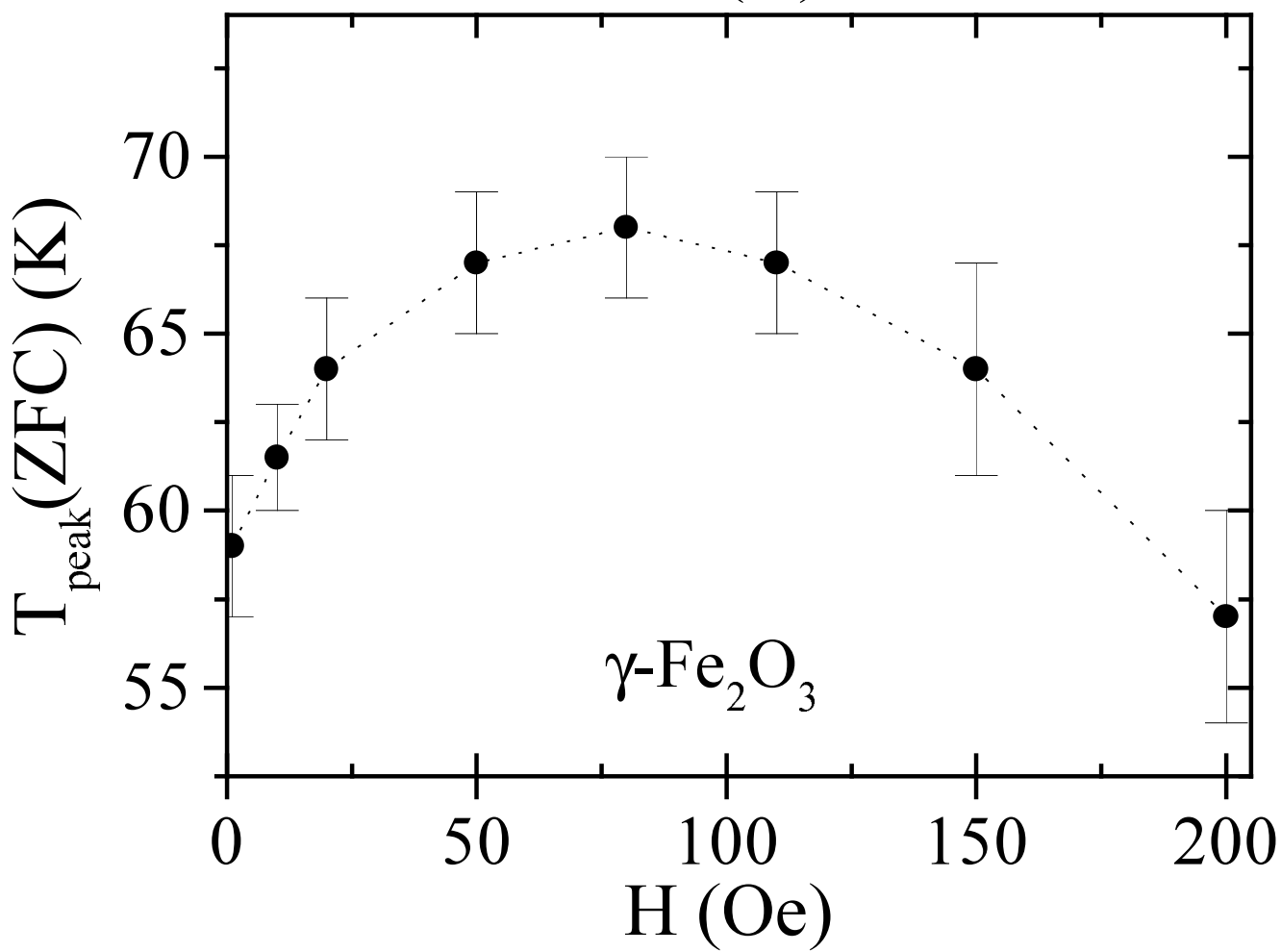
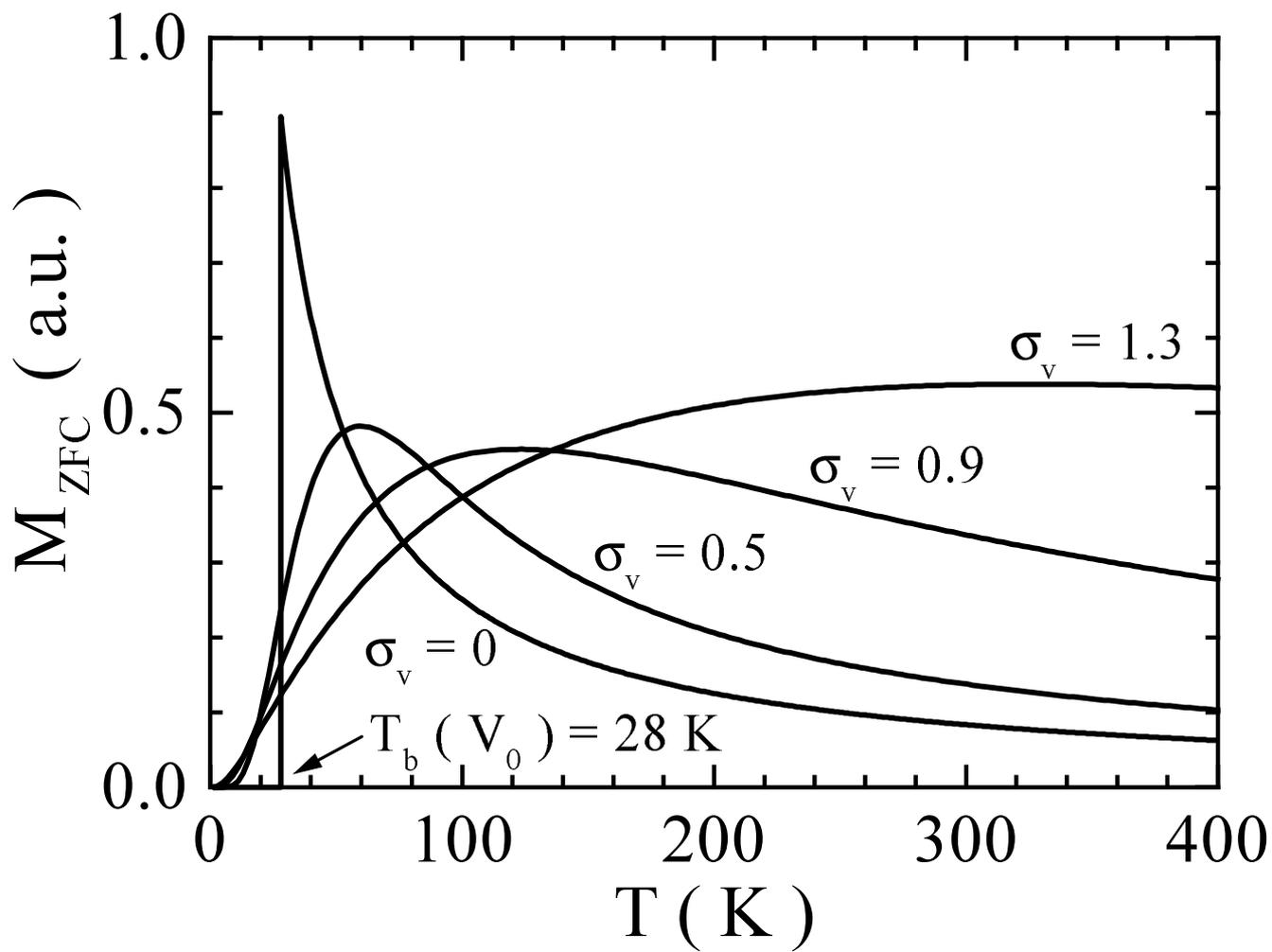
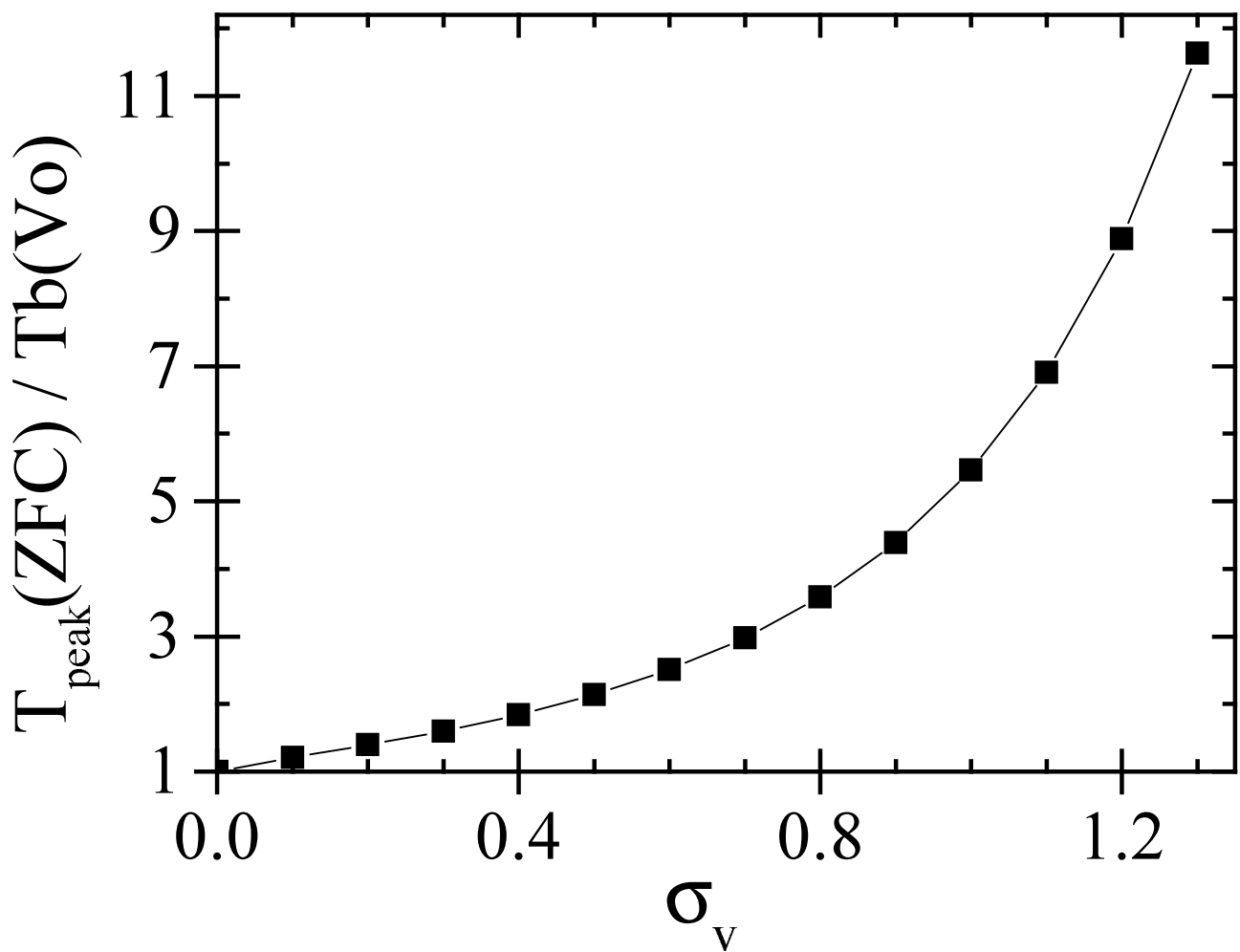
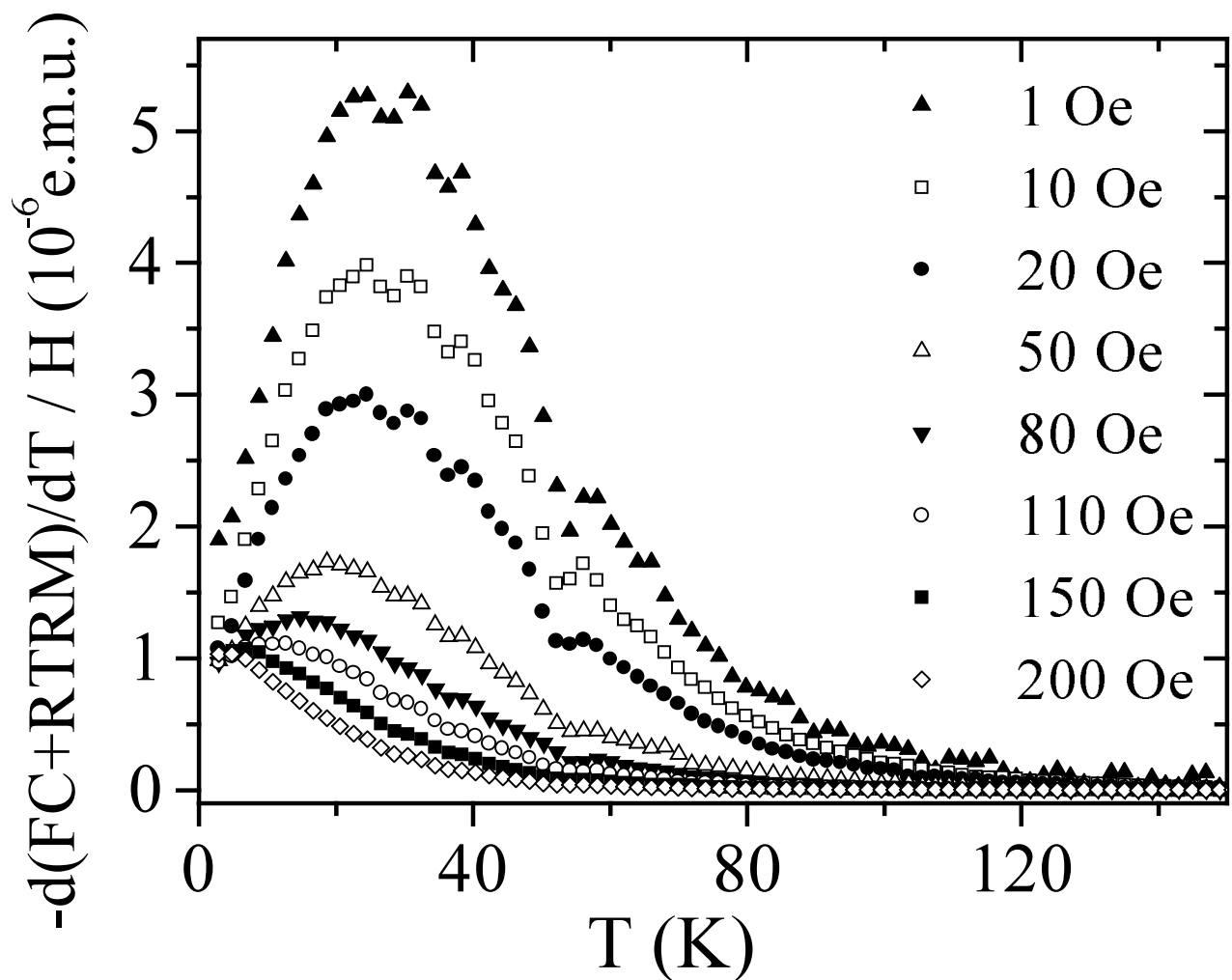
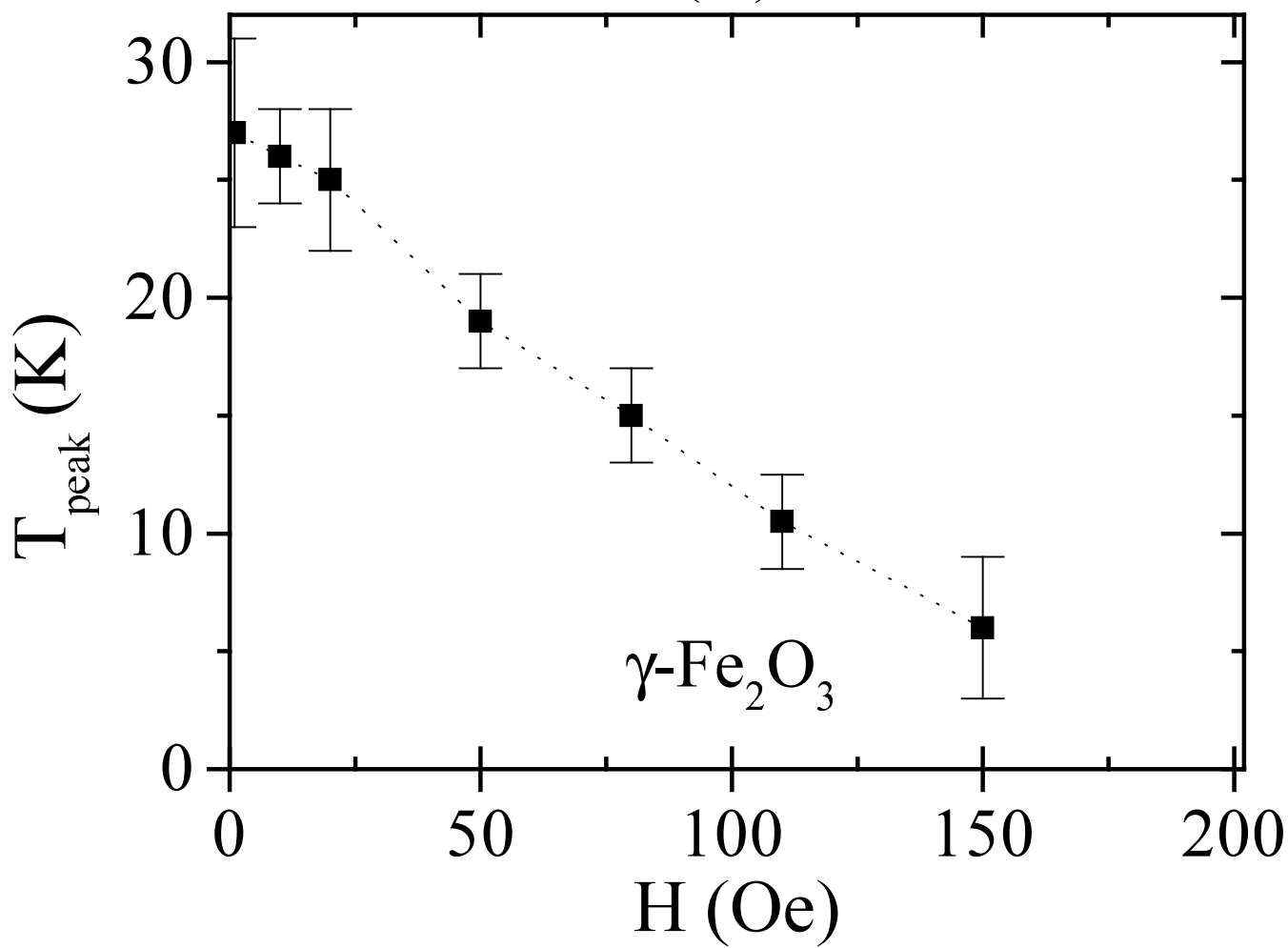
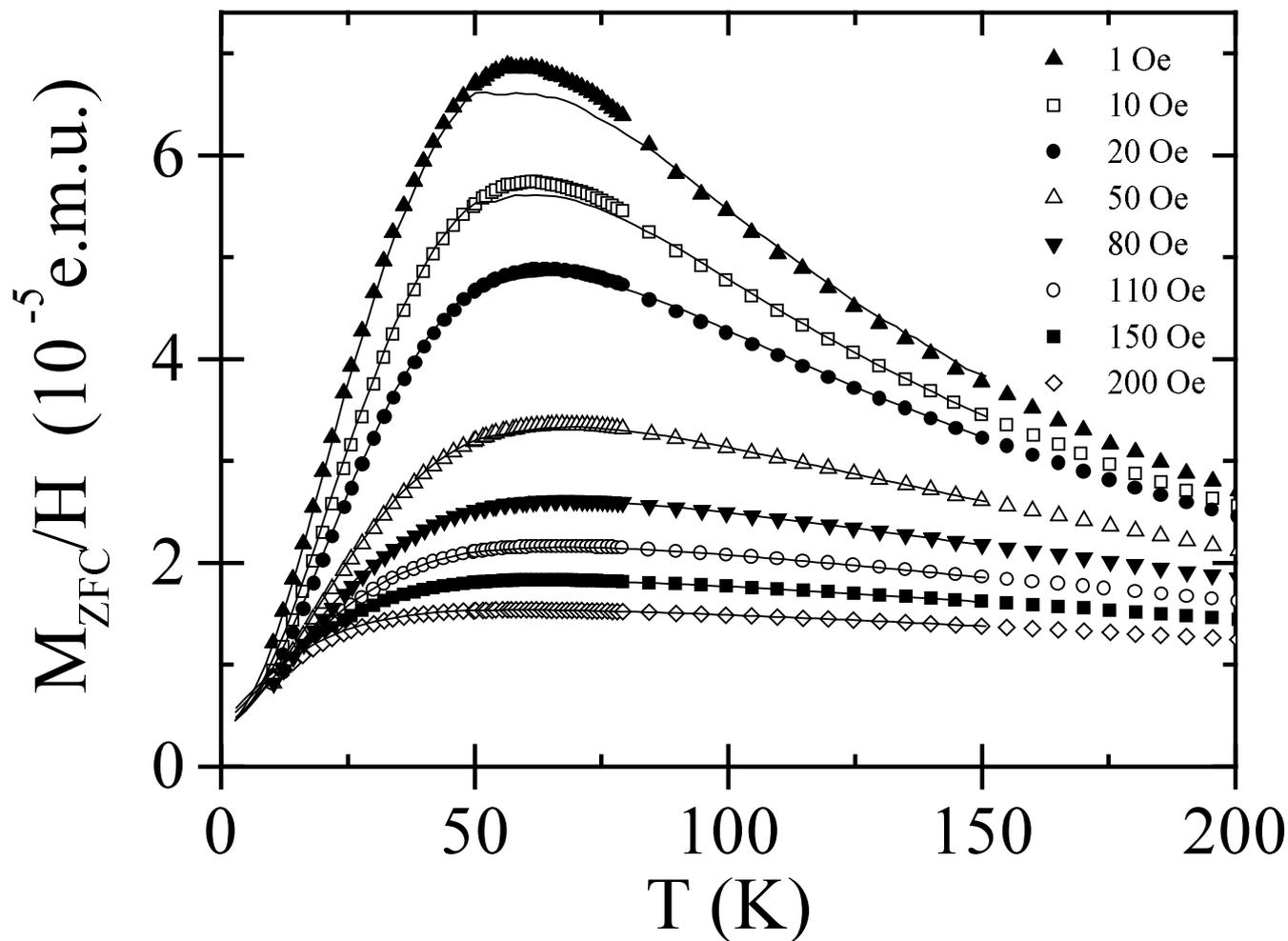
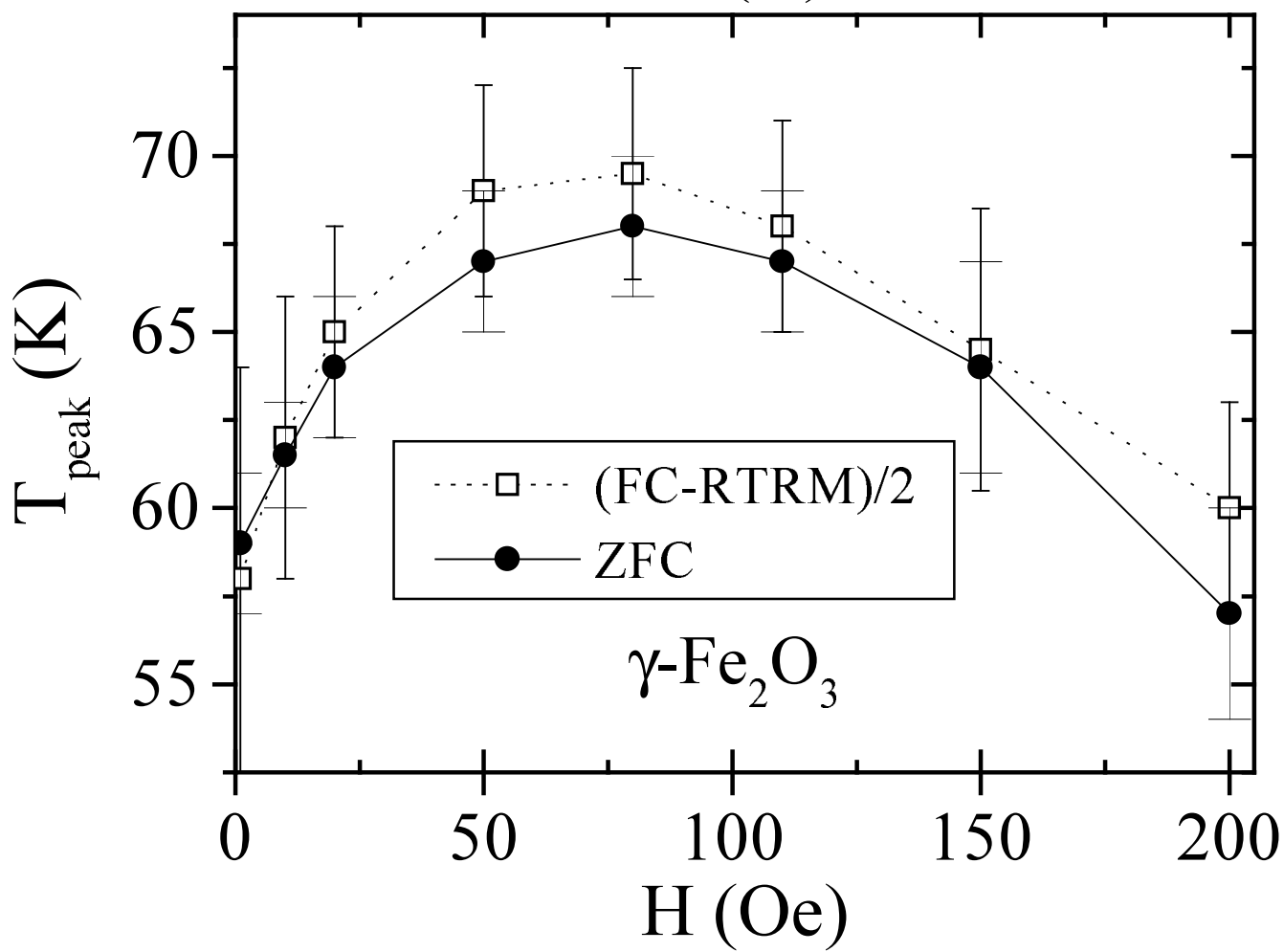
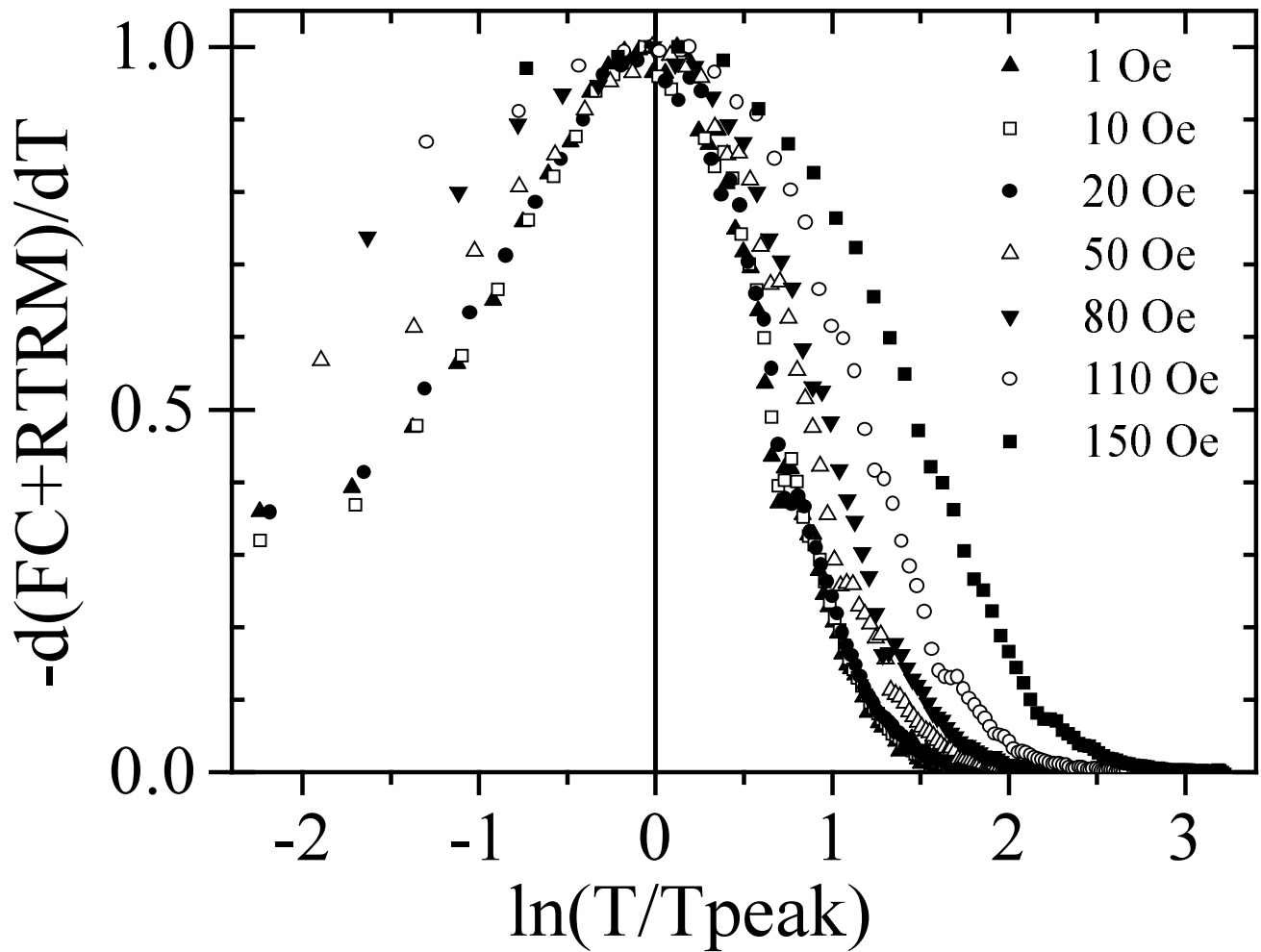
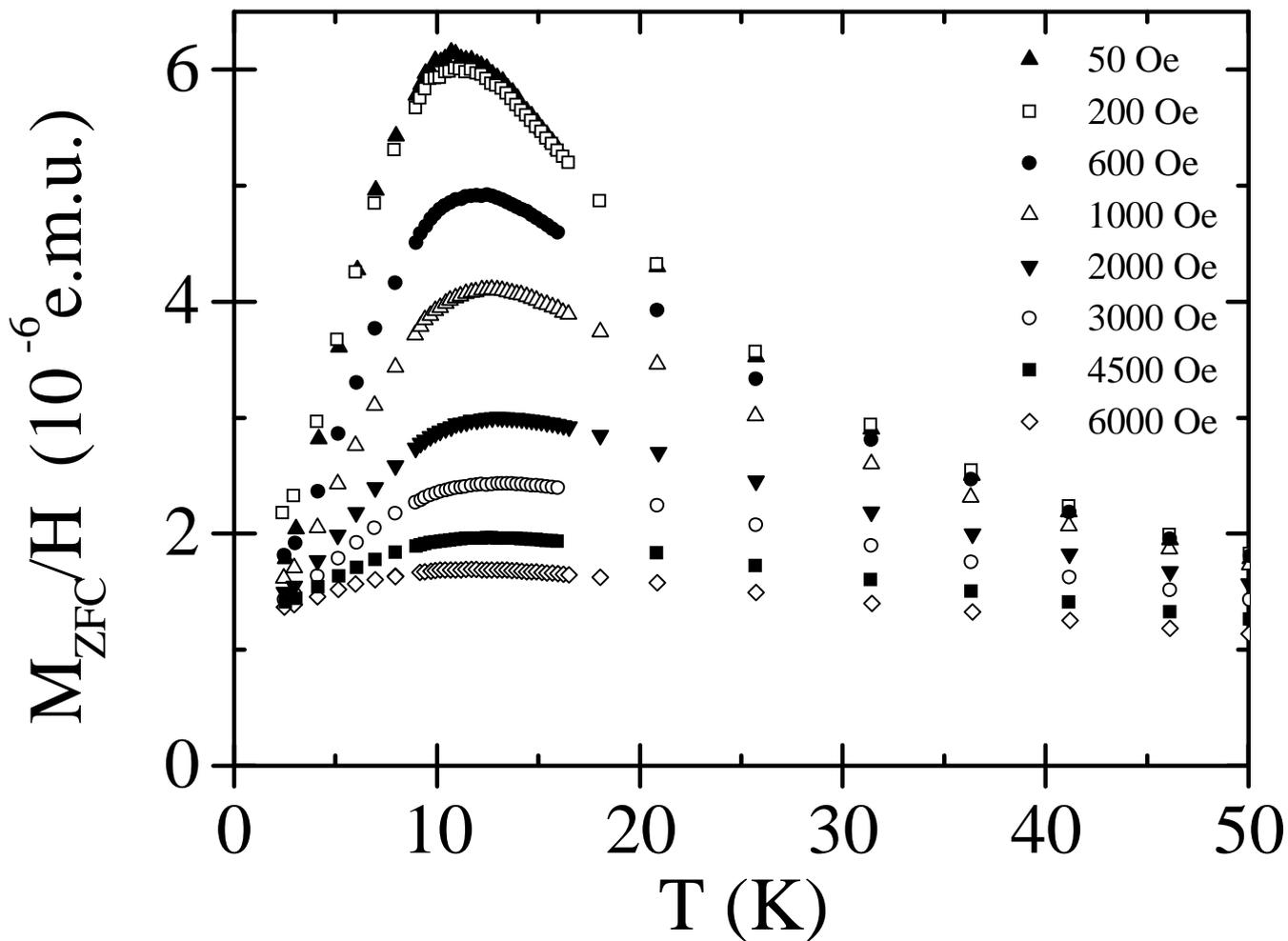
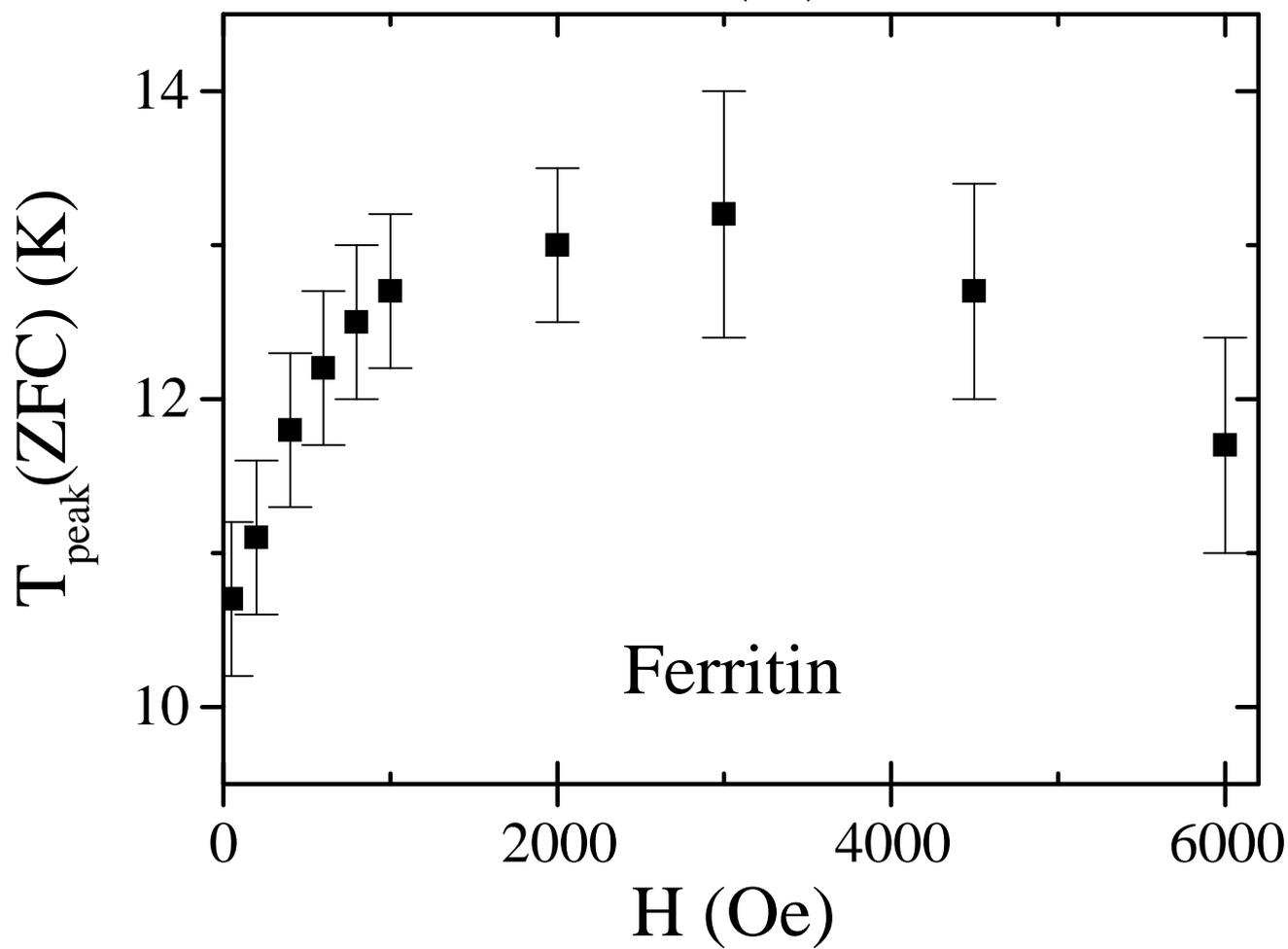
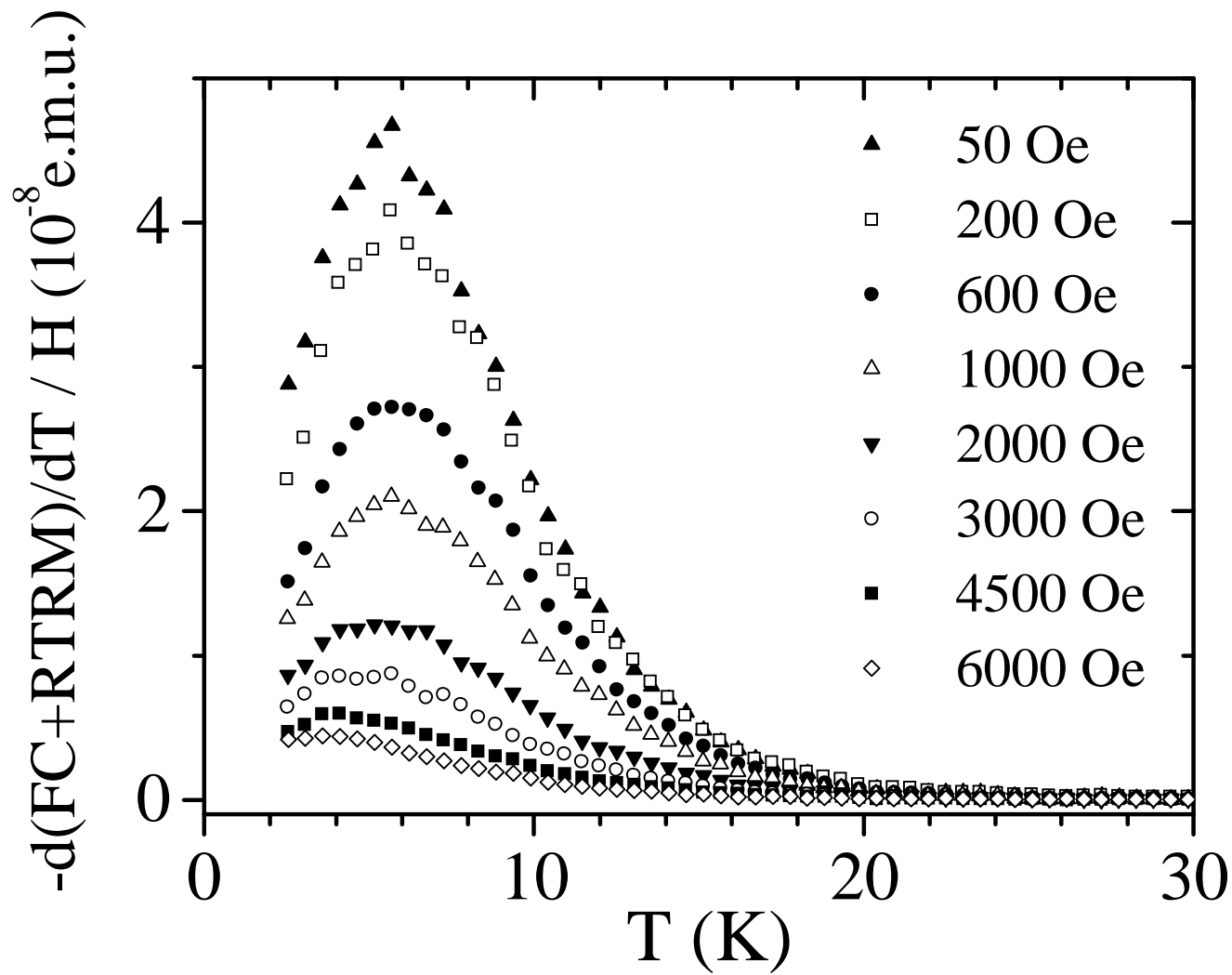
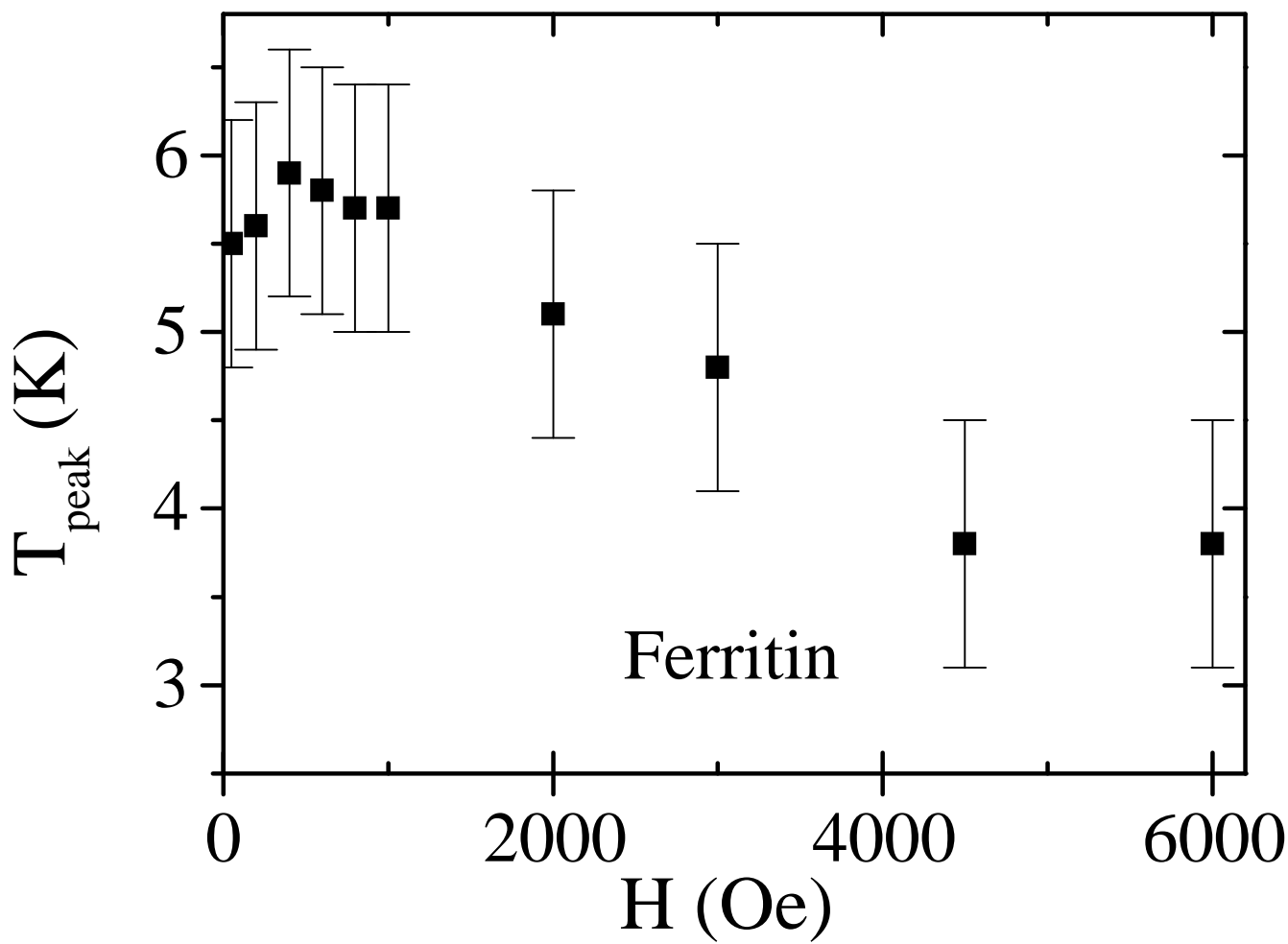
\end{document}